\documentclass[11pt]{article}

\usepackage{amsmath}
\usepackage{graphicx}
\usepackage{indentfirst}
\usepackage{amssymb}
\usepackage{cite}
\usepackage{color}
\usepackage{subfigure}

\begin{document}

\title{\bf{Quasinormal Modes of Massive Scalar Field with\\ Nonminimal Coupling in Higher-Dimensional de Sitter Black Hole with Single Rotation}}

\date{}
\maketitle

\begin{center}
\author{Bogeun Gwak}$^a$\footnote{rasenis@dgu.ac.kr}\\

\vskip 0.25in
$^{a}$\it{Division of Physics and Semiconductor Science, Dongguk University, Seoul 04620,\\Republic of Korea}\\
\end{center}
\vskip 0.6in

{\abstract
{We analytically investigate the quasinormal modes of the massive scalar field with a nonminimal coupling in the higher-dimensional de Sitter black hole with a single rotation. According to the separated scalar field equation, the boundary conditions of quasinormal modes are well constructed at the outer and cosmological horizons. Then, under near-extremal conditions, where the outer horizon closes to the cosmological horizon, the quasinormal frequencies are obtained and generalized to universal form in the higher-dimensional spacetime. Here, the real part of the frequency includes the scalar field contents, and its imaginary part only depends on the surface gravity at the outer horizon of the black hole.}}

\thispagestyle{empty}
\newpage
\setcounter{page}{1}

\section{Introduction}\label{sec1}

Black holes are compact objects in which a curvature singularity is located. The singularity is hidden by an event horizon, so the outside observer cannot see it. This plays an important role in black hole physics because the causal structure of spacetime is broken down, and physics is no longer predictive under a naked singularity without the horizon. Cosmic censorship was conjectured to avoid the breakdown of causality\cite{Penrose:1964wq, Penrose:1969pc, Hawking:1969sw}. There are two forms of the cosmic censorship conjecture: weak and strong. The weak cosmic censorship (WCC) conjecture states that the singularity should be invisible to an asymptotic observer owing to the horizon covering it. Hence, in the WCC conjecture, as long as the horizon is stable against perturbation, the singularity is still invisible. Then, the cosmic censorship is valid. It should be noted that there is no general proof for the conjecture, so its validity should be verified in each case. The first investigation of the WCC conjecture was conducted on the Kerr black hole by adding a particle, and it was found to be valid\cite{Wald:1974ge}. Furthermore, the validity depends on the states of a black hole as well as perturbation channels; the horizon of the near-extremal Kerr black hole becomes unstable owing to over-spinning by a particle\cite{Jacobson:2009kt}, but it is found to be stable when a self-force effect is introduced\cite{Barausse:2010ka,Barausse:2011vx,Colleoni:2015ena,Colleoni:2015afa,Sorce:2017dst}. Up to now, various studies on the WCC conjecture have been conducted for a diverse range of black holes, including cases with a cosmological constant\cite{Hubeny:1998ga, Crisostomo:2003xz, Isoyama:2011ea,Gao:2012ca,Rocha:2014jma,Cardoso:2015xtj,Gwak:2017kkt,Revelar:2017sem,An:2017phb,Ge:2017vun,Yu:2018eqq,Han:2019kjr}. Particularly, perturbation of a black hole owing to an external field is closely related to the stability of the horizon, so the WCC conjecture can be verified under the perturbation\cite{Hod:2008zza,Semiz:2005gs,Toth:2011ab,Duztas:2013wua,Natario:2016bay,Gwak:2018akg,Duztas:2018adf,Gwak:2019asi,Chen:2019nsr,Gwak:2019rcz}.

The perturbation of a black hole is more important in the strong cosmic censorship (SCC) conjecture than in the WCC conjecture. The SCC conjecture states that the singularity should also be hidden to an infalling observer as well as an asymptotic one. Here, the spacelike singularity is unimportant because the infalling observer cannot escape to the outside of the black hole. It should be noted that the timelike singularity is different, and an observer can see it. This can be a problem to the SCC conjecture. However, the Cauchy (or inner) horizon, enclosing the timelike singularity, can be a null singularity, when it interacts with the back-reaction\cite{Matzner:1979zz,Poisson:1990eh,Ori:1991zz}. Further, there is a possibility that the timelike singularity inside the null singularity (Cauchy horizon) becomes a spacelike singularity \cite{Brady:1995ni}. This ensures the SCC conjecture for asymptotically flat black holes. This situation becomes more complicated in de Sitter (dS) spacetime. Away from the cosmological horizon, the wave is redshifted, so perturbation by the field could make the Cauchy horizon stable. Hence, the validity of the SCC conjecture is closely related to how much the ingoing wave is efficiently damped around the outer horizon\cite{Brady:1998au}, and this can be clarified by analysis of the quasinormal modes (QNMs) in black holes. Further, more specifically, the QNMs should not be extendible across the Cauchy horizon and locally square integrable derivatives\cite{Christodoulou:2008nj}. The SCC conjecture is well studied in the Reissner--Nordstr\"{o}m-de Sitter (RNdS) black hole. The Cauchy horizon is known to be unstable to a scalar perturbation\cite{Brady:1998au}, but the QNMs in RNdS black holes were recently divided into three families, among which two are not sufficiently damped in near-extremal conditions. Hence, this makes the Cauchy horizon stable\cite{Cardoso:2017soq}. This violation of the SCC conjecture still remains in QNMs of the charged scalar field \cite{Dias:2018ufh}. To clarify the SCC conjecture in RNdS black holes, studies are currently being conducted from a variety of perspectives\cite{Burikham:2017gdm,Luna:2018jfk,Ge:2018vjq,Destounis:2018qnb,Gwak:2018rba,Hod:2018dpx,Cardoso:2018nvb,Dias:2018etb,Mo:2018nnu,Tangphati:2018jdx,Gim:2019rkl,Liu:2019lon,Liu:2019rbq,Zhang:2019ynp,Destounis:2019omd,Gan:2019jac,Guo:2019tjy,Sporea:2019iwk}. The investigation of the SCC conjecture is limited to rotating black holes\cite{Hod:2018lmi,Rahman:2018oso}.

In this work, we investigate the QNMs of the massive scalar field with a nonminimal coupling in the higher-dimensional dS black hole with a single rotation. In higher dimensions, the balance between gravity and centrifugal force differs from four-dimensional cases, so the properties of black holes are also different. Between them, the case with a single rotation is distinct in the asymptotically flat geometry over five dimensions because there is no upper bound on the spin parameter. Again, such Kerr-extremal conditions do not exist herein. Interestingly, due to the asymptotically dS geometry, the single rotation black hole has Nariai-type extremal conditions where outer and cosmological horizons are coincident. Even in black holes with a single rotation, Nariai-type extremal conditions are inevitable in any number of dimensions, and this is only seen in the dS black hole with a single rotation. Here, in consideration of the QNMs of the massive scalar field with a nonminimal coupling, we obtain the general form of the quasinormal frequencies in four and higher dimensions under near Nariai-type extremal conditions. Moreover, the quasinormal frequencies of the massive scalar field with a nonminimal coupling can be only analyzed by this field equation because there are no correspondences to the mass and nonminimal coupling with respect to null geodesic orbits. Particularly, the mass and nonminimal coupling of the scalar field are obtained in the real part of the frequency, and the imaginary part is only related to the surface gravity of the black hole. This implies that the decay rate of the QNMs only depends on the surface gravity in any dimension. The decay rate is sufficient to ensure the SCC conjecture, so we can expect this value. However, in higher dimensions with a single rotation, only an outer horizon and a cosmological horizon exist; therefore, the singularity is spacelike. Furthermore, the decay rate is still obtained in the universal form.

The remainder of this paper is organized as follows: In Section\,\ref{sec2}, we introduce the higher-dimensional dS black hole with a single rotation. In Section\,\ref{sec3}, the scalar field equation is solved at the outer horizon, and the boundary condition for QNMs is clarified. In Section\,\ref{sec4}, the quasinormal frequencies are obtained under near Nariai-type extremal conditions. In Section\,\ref{sec5}, we discuss our results in terms of the SCC conjecture. Finally, Section\,\ref{sec6} summarizes our results.

\section{Higher-Dimensional de Sitter Black Hole with Single Rotation}\label{sec2}

Here, we consider higher-dimensional rotating dS black holes in Einstein's gravity. The general form of a black hole is given in \cite{Gibbons:2004js}, where the black hole has many angular momenta corresponding to each rotation plane. Here, we assume that the black hole has only a single rotation. The solution to this was given in an earlier work \cite{Hawking:1998kw}. The metric of the black hole with a single rotation is given in the $D$-dimensional spacetime as
\begin{align}\label{eq:metric01}
ds^2&=-\frac{\Delta_r}{\rho^2}\left(dt-\frac{a\sin^2\theta}{\Xi} d\phi\right)^2+\frac{\rho^2}{\Delta_r}dr^2+\frac{\rho^2}{\Delta_\theta}d\theta^2+\frac{\Delta_\theta\sin^2\theta}{\rho^2}\left(adt-\frac{r^2+a^2}{\Xi}d\phi\right)^2+r^2 \cos^2\theta d\Omega_{D-4},\nonumber\\\rho^2&=r^2+a^2\cos^2\theta,\quad\Delta_r=(r^2+a^2)(1-\frac{\Lambda}{3} r^2)-\frac{2M}{r^{D-5}},\quad\Delta_\theta=1+\frac{\Lambda}{3}a^2\cos^2\theta,\quad\Xi=1+\frac{\Lambda}{3}a^2,
\end{align}
where $M$ and $a$ are mass and spin parameters, respectively. The metric in Eq.\,(\ref{eq:metric01}) represents $D$ dimensions in two parts. One is the four dimensions where the rotation plane exists. The other is the $(D-4)$-dimensional sphere perpendicular to the other four dimensions. The form of the $(D-4)$-dimensional sphere is given as
\begin{align}
d\Omega_{D-4}=\sum^{D-4}_{i=1}\left( \prod_{j=1}^{i} \sin^2\psi_{j-1}\right)d\psi_i^2,\quad \psi_0\equiv\frac{\pi}{2},\quad \Omega_{D-2}=\frac{2\pi^\frac{D-1}{2}}{\Gamma(\frac{D-1}{2})}.
\end{align}
The mass and angular momentum of the black hole in Eq.\,(\ref{eq:metric01}) are associated with the mass and spin parameters\cite{Altamirano:2013ane}.
\begin{align}\label{eq:massandangular}
M=\frac{\Omega_{D-2}}{4\pi}\frac{M}{\Xi^2}\left(1+\frac{(D-4)\Xi}{2}\right),\quad J=\frac{\Omega_{D-2}}{4\pi} \frac{Ma}{\Xi^2}.
\end{align}
The angular velocity and surface area at the outer horizon $r_\text{h}$ are 
\begin{align}\label{eq:area01}
\Omega_\text{rot}=\frac{a\Xi}{r_\text{h}^2+a^2},\quad \mathcal{A}=\frac{\Omega_{D-2}(r_\text{h}^2+a^2)r_\text{h}^{D-4}}{\Xi}.
\end{align}
However, the angular velocity does not tend to zero at the asymptotic boundary of the negative cosmological constant, which is an anti-de Sitter case. Instead, it is $-a/\ell^2$. This implies that the observer is rotating with respect to the boundary. Then, the effect of this rotation can be included in the energy of such a scalar field, so we cannot exactly define it. To resolve this rotating boundary, the static observer is introduced by the following coordinate transformation\cite{Hawking:1998kw}:
\begin{align}\label{eq:coordinatetransformation}
t\rightarrow T,\quad \phi\rightarrow \Phi+\frac{1}{3}a\Lambda T.
\end{align}
This transformation makes the angular velocity at the AdS boundary zero. Even if we consider only dS black holes, the effect of this asymptotic rotation is still included in the coordinate system, as in the AdS case. Hence, we apply the transformation in Eq.\,(\ref{eq:coordinatetransformation}) to the metric in Eq.\,(\ref{eq:metric01}). The transformed metric is 
\begin{align}\label{eq:metric03}
ds^2=&-\frac{\Delta_r}{\rho^2\Xi^2}\left(\Delta_\theta dT-a\sin^2\theta d\Phi\right)^2+\frac{\rho^2}{\Delta_r}dr^2+\frac{\Delta_\theta\sin^2\theta}{\rho^2\Xi^2}\left(a\left(1-\frac{\Lambda}{3} r^2\right)dT-(r^2+a^2)d\Phi\right)^2\\
&+\frac{\rho^2}{\Delta_\theta}d\theta^2+r^2\cos^2\theta d\Omega_{D-4},\nonumber
\end{align}
where mass and angular momentum are still defined as in Eq.\,(\ref{eq:massandangular}). Then, the angular velocity at the outer horizon is given by
\begin{align}
\Omega_\text{h}=\frac{a\left(1-\frac{\Lambda}{3}r_\text{h}^2\right)}{r_\text{h}^2+a^2},
\end{align}
which can be the true angular velocity measured by a static observer. The surface gravities at the outer and cosmological horizons, $r_\text{h}$ and $r_\text{c}$, are
\begin{align}\label{eq:ent07}
\kappa_\text{h}&=\frac{r_\text{h}\left((7-D)+(5- D)\frac{a^2}{r_\text{h}^2}-\frac{1}{3}(7-D)a^2 \Lambda-3(1-\frac{1}{3}D)r_\text{h}^2\Lambda\right)}{2\left(r_\text{h}^2+a^2\right)},\\ \kappa_\text{c}&=\frac{r_\text{c}\left((7-D)+(5- D)\frac{a^2}{r_\text{c}^2}-\frac{1}{3}(7-D)a^2 \Lambda-3(1-\frac{1}{3}D)r_\text{c}^2\Lambda\right)}{2\left(r_\text{c}^2+a^2\right)}.\nonumber
\end{align}
In a dS black hole with a single rotation, the outer horizon always exists for any value of the spin parameter in six or more dimensions, i.e., for $D\geq 6$. This is due to the form of the function $\Delta_r$ in Eq.\,(\ref{eq:metric01}). For a given value of $\Lambda$, in the ultra-spinning case, $a\gg 1$, the outer horizon is located at $r_\text{h}\sim \frac{1}{a^2}$. Further, this implies that the surface area of the black hole is shrunk due to Eq.\,(\ref{eq:area01}). However, in terms of the surface areas of the two-dimensional sphere in four dimensions and the $(D-4)$-dimensional sphere in additional spatial dimensions, the situation becomes quite different, because\cite{Emparan:2003sy}
\begin{align}
\mathcal{A}_2=\frac{\Omega_2 (r_\text{h}^2+a^2)}{\Xi}\simeq \frac{3}{\Lambda},\quad \mathcal{A}_{D-4} =\Omega_{D-4} (r_\text{h} \cos\theta)^{D-4}\simeq \left(\frac{1}{a^2}\right)^{D-4}\ll 1.
\end{align}
Thus, in the ultra-spinning case, the surface of the black hole does not shrink in four dimensions where the rotation plane is located, but the surface shrinks in the additional $(D-4)$ dimensions. This can be only observed in the case of a single rotation. For multiple rotations or the AdS cases, there are Kerr-extremal values of the spin parameters, so this kind of shrinking cannot be observed.

\section{Boundary Conditions for QNMs}\label{sec3}

Scattered by the rotating dS black hole, the external scalar field is assumed to be massive and nonminimally coupled to the curvature in $D$-dimensional spacetime. Here, we consider the QNMs of the scalar field obtained by imposing a specific boundary condition. The action of the scalar field is
\begin{align}\label{eq:fieldaction01}
S_\Psi =-\frac{1}{2}\int d^D x \sqrt{-g}\left(\partial_\mu \Psi \partial^\mu \Psi^*+(\mu^2+\xi \mathcal{R})\Psi\Psi^*\right),
\end{align}
where the mass and nonminimal coupling are denoted as $\mu$ and $\xi$, respectively. The curvature of the dS spacetime is given as a constant $\mathcal{R}={2(D-1)}/{(D-3)\Lambda}$. Hence, it acts as an effective mass of the scalar field but depends on the dimensionality. The field equation is derived from Eq.\,(\ref{eq:fieldaction01}) as
\begin{align}\label{eq:fieldeq01}
\frac{1}{\sqrt{-g}} \partial_\mu \left(\sqrt{-g} g^{\mu\nu} \partial _\nu \Psi\right)-(\mu^2+\xi \mathcal{R}) \Psi=0,\quad  \sqrt{-g}=\frac{\rho^2 \sin\theta}{\Xi}(r\cos\theta)^{D-4}\prod^{D-5}_{j=1}(\sin\psi_j)^{D-4-j}.
\end{align}
The field equation is separable, so solutions can take the form 
\begin{align}\label{eq:fieldeqsol01}
\Psi(T,r,\theta,\Phi)=e^{-i\omega T}e^{im\Phi} R(r) \Theta(\theta) Y_{\ell m'}(\psi_1,...\psi_{D-4}),
\end{align}
where the $(D-2)$-dimensional sphere of the form is divided into two parts: $(\theta,\phi)$ and $(\psi_1,...,\psi_{D-4})$. This originates from the metric represented in Eq.\,(\ref{eq:metric03}). Then, the radial equation is obtained from Eqs.\,(\ref{eq:fieldeq01}) and (\ref{eq:fieldeqsol01}) as
\begin{align}\label{eq:radialeq02}
\frac{1}{r^{D-4} R(r)} \partial_r\left(r^{D-4} \Delta_r \partial_r R(r)\right)+\frac{(r^2+a^2)^2}{\Delta_r} \left(\omega -m\Omega_\text{h}\right)^2-\frac{\ell(\ell+D-5)a^2}{r^2}-(\mu^2+\xi \mathcal{R})r^2-\lambda=0,
\end{align}
where the eigenvalue $\ell$ is about the angular momentum in the $(D-4)$-dimensional sphere, and the eigenvalue $\lambda$ is about the $2$-dimensional sphere. Further, the $\theta$-directional equation is also separated into
\begin{align}
&\frac{1}{\sin\theta \cos^{D-4}\theta\,\Theta(\theta)}\partial_\theta\left(\sin\theta \cos^{D-4}\theta \Delta_\theta \partial_\theta \Theta(\theta)\right)\\
&+\left(\lambda-\frac{a^2\omega^2 \sin^2\theta}{\Delta_\theta}-\frac{m^2\Delta_\theta }{\sin^2\theta}-\frac{\ell(\ell+D-5)}{\cos^2\theta}-(\mu^2+\xi \mathcal{R})a^2 \cos^2\theta+2\omega a m\right)=0.\nonumber
\end{align}
This is the generalized scalar hyperspheroidal equation, and its details are discussed in \cite{Berti:2005gp,Cho:2009wf}. In the limit of slow rotation, $a\omega\rightarrow 0$, the eigenvalue $\lambda$ goes to that of the hyperspherical harmonics on the $(D-2)$-dimensional sphere\cite{Gwak:2019asi}. Hence, its eigenvalue becomes $\ell (\ell+1)$ in the case without rotation. With a large value of rotation, the eigenvalue is quite different compared with the slowly rotating case and should be analyzed in each case\cite{Cho:2009wf}. Fortunately, our focus is on the quasinormal frequencies, which will be obtained from the radial equation of Eq.\,(\ref{eq:radialeq02}), so we simply take eigenvalues of the generalized scalar hyperspheroidal equation.

Here, in Eq.\,(\ref{eq:radialeq02}), we must transform the radial equation into a Schr\"{o}dinger-like equation using particular tortoise coordinate and form of the solution.
\begin{align}
\frac{dr^*}{dr}=\frac{(r^2+a^2)}{r^{D-4}\Delta_r},\quad R(r)\rightarrow \frac{R(r)}{\sqrt{r^2+a^2}},
\end{align}
where $r\rightarrow r_\text{h}$ and $r\rightarrow r_\text{c}$ are transformed into $r^*\rightarrow -\infty$ and $r^*\rightarrow +\infty$ under the tortoise coordinate, respectively. This changes the radial equation where there is no first derivative term. Thus, the radial equation becomes a Schr\"{o}dinger-like equation.
\begin{align}\label{eq:radialschrodinger01}
& \frac{(r^2+a^2)^2}{r^{2D-8}\Delta_r R}\frac{d^2 R }{dr^{*2}}-\frac{d\Delta_r r (r^2+a^2)+((D-3)a^2+(D-6)r^2)\Delta_r}{(r^2+a^2)^2}\\
&+\frac{(r^2+a^2)^2}{\Delta_r} \left(\omega -m\Omega_\text{h}\right)^2-\frac{\ell(\ell+D-5)a^2}{r^2}-(\mu^2+\xi \mathcal{R})r^2-\lambda=0,\nonumber
\end{align}
where
\begin{align}
d\Delta_r = \frac{d \Delta _r}{d r }= 2r\left(1-\frac{\Lambda}{3}r^2\right)-\frac{2}{3}r (r^2+a^2)+2(D-5)\frac{M}{r^{D-4}}.
\end{align}
In the dS black hole, there are two boundaries: $r\rightarrow r_\text{h}$ and $r\rightarrow r_\text{c}$. At these boundaries, the radial equation in Eq.\,(\ref{eq:radialschrodinger01}) becomes
\begin{align}\label{eq:boundaryeq02}
\frac{d^2 R }{dr^{*2}}+(r_\text{h}^{D-4})^2\left(\omega -m\Omega_\text{h}\right)^2R=0,\quad \frac{d^2 R }{dr^{*2}}+(r_\text{c}^{D-4})^2\left(\omega -m\Omega_\text{c}\right)^2R=0.
\end{align}
Satisfying Eq.\,(\ref{eq:boundaryeq02}), the radial solutions are at the outer horizon
\begin{align}\label{eq:horizonsol01}
R(r)\sim e^{\pm i r_\text{h}^{D-4}(\omega - m \Omega_\text{h})r^*},
\end{align} 
and the radial solutions are at the cosmological horizon
\begin{align}\label{eq:horizonsol02}
R(r)\sim e^{\pm i r_\text{c}^{D-4}(\omega - m \Omega_\text{c})r^*}.
\end{align}
In combination with Eqs.\,(\ref{eq:horizonsol01}) and (\ref{eq:horizonsol02}), we can define the boundary conditions for the scalar field. Particularly, the QNMs require purely ingoing waves at the outer horizon and purely outgoing waves at the cosmological horizon\cite{Chambers:1997ef}. Hence, the boundary conditions for QNMs are
\begin{align}\label{eq:boundarycondition03a}
R(r)\sim e^{-i r_\text{h}^{D-4}(\omega - m \Omega_\text{h})r^*},\quad r^* \rightarrow -\infty;\quad
R(r)\sim e^{+ i r_\text{c}^{D-4}(\omega - m \Omega_\text{c})r^*},\quad r^* \rightarrow +\infty.
\end{align}
We ensure that these boundary conditions satisfy the field equation in Eq.\,(\ref{eq:boundaryeq02}). Then, under the boundary conditions, the QNMs are obtained.

\section{QNMs in Near Nariai-Type Extremal Black Holes}\label{sec4}

QNM in higher-dimensional dS black holes was studied in the Schwarzschild--dS black hole\cite{Konoplya:2003dd,Konoplya:2007jv}. Because rotating black holes are physically rich, QNMs in the Kerr--dS black hole have been investigated considering various perspectives\cite{Konoplya:2007zx,Dyatlov:2010hq,Dyatlov:2011jd,Kraniotis:2016maw,Novaes:2018fry}. Furthermore, in higher dimensions, the balance between gravity and centrifugal force differs from the four-dimensional case; therefore, the characteristics of QNMs can be different in higher-dimensional rotating-dS black holes\cite{Barragan-Amado:2018pxh}. Recently, the decay rate of the dominant QNM was found to be crucial in the SCC conjecture in Kerr--dS black holes\cite{Hod:2018lmi,Rahman:2018oso,Dias:2018ynt}.

Here, we analytically investigate the frequencies of the QNMs in near Nariai-type extremal black holes, where the location of the outer horizon closes to that of the cosmological horizon, so $r_\text{h}\approx r_\text{c}$.
\begin{figure}[h]
\centering
\subfigure[{Phase diagram in $D=4$.}] {\includegraphics[scale=0.9,keepaspectratio]{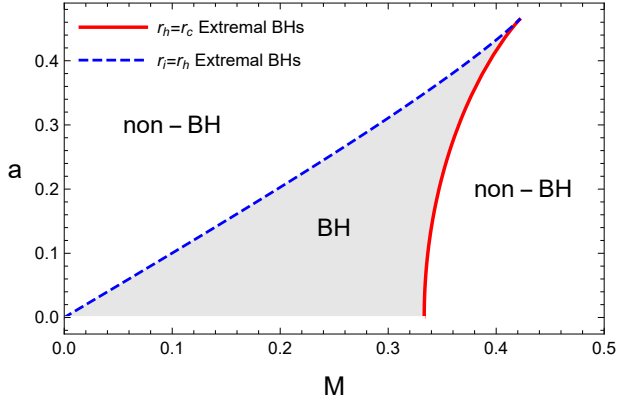}}\quad\quad
\subfigure[{Phase diagram in $D=5$.}] {\includegraphics[scale=0.9,keepaspectratio]{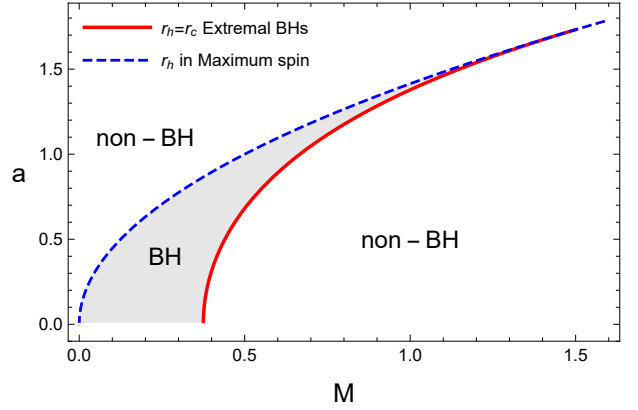}}\\
\subfigure[{Phase diagram in $D=6$.}] {\includegraphics[scale=0.9,keepaspectratio]{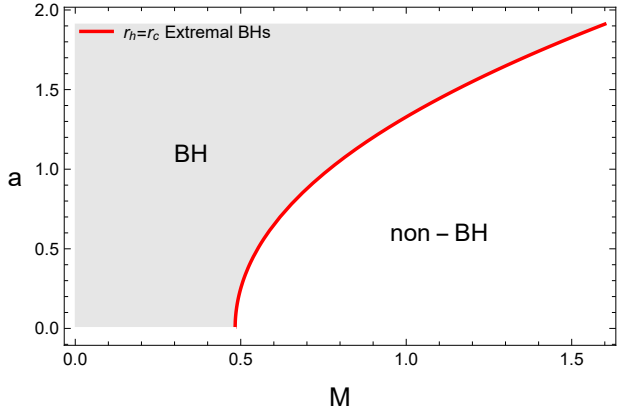}}\quad\quad
\subfigure[{Phase diagram in $D=7$.}] {\includegraphics[scale=0.9,keepaspectratio]{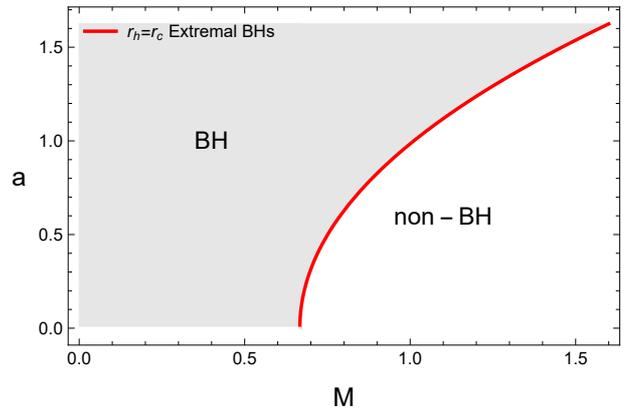}}
\caption{{\small Phase diagrams about mass and spin parameters for the positive cosmological constants, $\Lambda=1$.}}
\label{fig:fig1}
\end{figure}
Because the metric function $\Delta_r$ depends on the dimensionality, the solution spaces of the black hole are different in each dimension. Particularly, we can divide these into three cases owing to the power of gravity: $D=4$, $D=5$, and $D\geq 6$. Their phases are shown in Fig.\,\ref{fig:fig1}. Here, our interest is in the red lines around the near Nariai-type extremal black hole of $r_\text{h}\approx r_\text{c}$. In Fig.\,\ref{fig:fig1}\,(a), the four-dimensional black hole can represent two types of extremal black holes: $r_\text{i}=r_\text{h}$ and $r_\text{h}=r_\text{c}$. This is denoted by the gray area, and the other phases are not black holes. In Fig.\,\ref{fig:fig1}\,(b), there is no Kerr-extremal black hole of $r_\text{i}=r_\text{h}$. Instead, there is a maximum spin parameter where the black hole can exist. However, in Fig.\,\ref{fig:fig1}\,(c) and (d), for $D\geq 6$, there are no limits on the spin parameters, so the Nariai-type extremal black holes only exist at $r_\text{h}=r_\text{c}$. According to these phase spaces, our analysis is commonly applicable to Nariai-type extremal black holes of $r_\text{h}=r_\text{c}$ in all dimensions denoted by the red lines.

The QNM frequencies are obtained from the Schr\"{o}dinger-like radial equation in Eq.\,(\ref{eq:radialschrodinger01}) with near Nariai-type extremal conditions. Under the near Nariai-type extremal conditions, the metric function $\Delta_r$ satisfies
\begin{align}
\Delta_r\ll 1,\quad d\Delta_r\ll 1.
\end{align}
Then, by taking the leading term, the radial equation is simplified as
\begin{align}
\frac{d^2 R }{dr^{*2}}+\left(\omega r^{D-4} -m\Omega_\text{h} r^{D-4}\right)^2R+\frac{r^{2D-8}\Delta_r }{(r^2+a^2)^2}\left(-\frac{\ell(\ell+D-5)a^2}{r^2}-(\mu^2+\xi \mathcal{R})r^2-\lambda\right)R=0,
\end{align}
where the radial coordinate is $r\approx r_\text{h}\approx r_\text{c}$ owing to the near Nariai-type extremal conditions. Further, in the near Nariai-type extremal black hole, the metric function $\Delta_r$ can be rewritten in decomposed form\cite{Cardoso:2003sw,Molina:2003ff}. Then,
\begin{align}
\Delta_r = \frac{3}{\Lambda}\frac{1}{r^{D-5}} (r-r_\text{h})(r_\text{c}-r)(r-r_1)...(r-r_{D-3}),\quad d\Delta_r \approx \frac{3}{\Lambda}\frac{1}{ r^{D-5}} (r_\text{c}-r_\text{h})(r_\text{h}-r_1)...(r_\text{h}-r_{D-3}).
\end{align}
This can be generalized to the rotating black hole as
\begin{align}\label{eq:nearextremaltortoise01}
\Delta_r&=\frac{d\Delta_r}{r_\text{c}-r_\text{h}}(r-r_\text{h})(r_\text{c}-r)=\frac{2\kappa_\text{h}(r_\text{h}^2+a^2)}{r_\text{c}-r_\text{h}}(r-r_\text{h})(r_\text{c}-r).
\end{align}
According to Eq.\,(\ref{eq:nearextremaltortoise01}), we can find an exact form of the tortoise coordinate in terms of $r$.
\begin{align}\label{eq:nearextremaltortoise02}
r^*=\frac{1}{2\kappa_\text{h} r_\text{h}^{D-4}}\left(\ln(r-r_\text{h})-\ln(r_\text{c}-r)\right),\quad r=\frac{r_\text{h}+r_\text{c} e^{2\kappa_\text{h} r_\text{h}^{D-4} r^*}}{1+e^{2\kappa_\text{h} r_\text{h}^{D-4} r^*}}.
\end{align}
By inserting Eq.\,(\ref{eq:nearextremaltortoise02}) into Eq.\,(\ref{eq:nearextremaltortoise01}), an approximation to the metric function $\Delta_r$ is obtained as
\begin{align}
\Delta_r=\frac{(r_\text{c}-r_\text{h})(r_\text{h}^2+a^2)\kappa_\text{h}}{2\cosh^2(\kappa_\text{h}r_\text{h}^{D-4}r^*)}.
\end{align}
Hence, for the near Nariai-type extremal black hole, the Schr\"{o}dinger-like radial equation becomes 
\begin{align}
\frac{d^2 R }{dr^{*2}}+\left(\left(\omega r_\text{h}^{D-4} -m\Omega_\text{h} r_\text{h}^{D-4}\right)^2-\frac{V_0}{\cosh^2(\kappa_\text{h}r_\text{h}^{D-4}r^*)}\right)R=0,
\end{align}
where
\begin{align}
V_0=\frac{r_\text{h}^{2D-8}}{(r_\text{h}^2+a^2)}\frac{(r_\text{c}-r_\text{h})\kappa_\text{h}}{2}\left(\frac{\ell(\ell+D-5)a^2}{r_\text{h}^2}+(\mu^2+\xi \mathcal{R})r_\text{h}^2+\lambda\right).
\end{align}
This is P\"{o}shl-Teller potential\cite{Poschl:1933zz}, which is solvable under the boundary conditions of quasinormal frequencies in Eq.\,(\ref{eq:boundarycondition03a})\cite{Ferrari:1984zz}. Then, we obtain the quasinormal frequencies, where
\begin{align}
\omega =m\Omega_\text{h} +\kappa_\text{h}\sqrt{\frac{V_0}{\kappa_\text{h}^2r_\text{h}^{2D-8}}-\frac{1}{4}}-i\left(n+\frac{1}{2}\right)\kappa_\text{h},\quad n=0,\,1,\,2,...
\end{align}
The quasinormal frequencies can be divided into their real and imaginary parts.
\begin{align}\label{eq:qnmsfresol01}
\text{Re}(\omega)&=m\Omega_\text{h} +\kappa_\text{h}\sqrt{\frac{r_\text{c}-r_\text{h}}{2\kappa_\text{h}(r_\text{h}^2+a^2)}\left(\frac{\ell(\ell+D-5)a^2}{r_\text{h}^2}+(\mu^2+\xi \mathcal{R})r_\text{h}^2+\lambda\right)-\frac{1}{4}},\\
\text{Im}(\omega)&=-\left(n+\frac{1}{2}\right)\kappa_\text{h}.\nonumber
\end{align}
Therefore, these are the quasinormal frequencies of the near Nariai-type extremal dS black hole with a single rotation, which implicitly depends on dimensionality. Further, we expect that Eq.\,(\ref{eq:qnmsfresol01}) is universally applicable to $D$-dimensional cases. According to the imaginary part in Eq.\,(\ref{eq:qnmsfresol01}), the most dominant mode of the QNMs is the $n=0$ case of the least damping mode. Note that the imaginary part also depends on the value of the square root of the real part in Eq.\,(\ref{eq:qnmsfresol01}); however, the absolute value of the imaginary part in a dominant quasinormal mode is smaller than $\frac{1}{2}\kappa_\text{h}$. Hence, this cannot change the physical implications considerably. Particularly, the decay rate of the massive scalar field with a nonminimal coupling is formulated in the same way regardless of the scalar field's characteristics, dimensionality, or phase structure of the solutions in the near Nariai-type extremal black hole.

\section{Implication in Strong Cosmic Censorship Conjecture}\label{sec5}

The imaginary part of the quasinormal frequency in Eq.\,(\ref{eq:qnmsfresol01}) is the decay rate of the nonminimally coupled massive scalar field in the Nariai-type extremal black hole in $D$ dimensions. Hence, the decay rate only depends on the integer $n$, which represents different modes. Among the modes, the dominant mode is the least damped one. Then, the decay rate of the dominant mode is
\begin{align}\label{eq:scc03}
|\text{Im}(\omega_{n=0})|=\frac{1}{2}\kappa_\text{h}.
\end{align} 
This is universal in $D$-dimensional dS black holes with a single rotation. The decay rate of the dominant QNM plays an important role in the SCC conjecture for black holes with an inner horizon. Here, according to Fig.\,\ref{fig:fig1} in the case of a single rotation, the four-dimensional Kerr-dS black hole is the only case that we can consider. In other dimensions, there only exist outer and cosmological horizons, so they are not applicable to the SCC conjecture.

In the four-dimensional case, the SCC conjecture depends on how much the scalar field efficiently decays in the black hole exterior compared with amplification in its interior\cite{Hintz:2015jkj}. Further, with the positive cosmological constant, the scalar field is exponentially decayed as
\begin{align}
|\Psi-\Psi_0|\leq A e^{-\alpha t},
\end{align}
where $\Psi_0$ is a constant, and $\alpha$ is the spectral gap determined by the decay rate of the dominant QNM. Then, the competition between decay and amplification can be denoted as $\beta\equiv \alpha / |\kappa_\text{i}|$. According to the value of $\beta$, the following behavior of the scalar field can be expected: if $\beta<1$, the scalar field cannot be extended across the Cauchy horizon. Moreover, for the validity of the SCC conjecture, the energy of the scalar field must diverge at the Cauchy horizon\cite{Christodoulou:2008nj}. Again, the scalar field should not have locally square integrable derivatives\cite{Dias:2018ynt,Dias:2018etb,Dias:2018ufh,Cardoso:2018nvb}. In terms of $\beta$, the following arises:
\begin{align}\label{eq:scc02}
\beta<\frac{1}{2}.
\end{align}
Note that because the nonminimal coupling and mass of the scalar field are considerably small values compared with the mass of the black hole, the square integrable condition in Eq.\,(\ref{eq:scc02}) corresponds to that of the Kerr-dS black hole\cite{Dias:2018ynt}. However, if the nonminimal coupling and mass of the scalar field is large compared with the mass of the black hole, the condition in Eq.\,(\ref{eq:scc02}) should be modified and different. According to Eq.\,(\ref{eq:scc02}), the decay rate of the dominant QNM in Eq.\,(\ref{eq:scc03}) can be rewritten as
\begin{align}
\beta = \frac{1}{2}\frac{|\text{Im}(\omega_{n=0})|}{|\kappa_\text{i}|}<\frac{1}{2},\quad (\because \kappa_\text{h} < |\kappa_\text{i}|).
\end{align}
Thus, our results support the SCC conjecture being valid in the near Nariai-type extremal Kerr--dS black hole under the massive scalar field with a nonminimal coupling. This is consistent with \cite{Dias:2018ynt} in the nonextremal Kerr--dS black hole with the massless scalar field, \cite{Hod:2018lmi,Rahman:2018oso} in the near-extremal Kerr--Newman--dS, and higher-dimensional rotating dS black holes by the Lyapunov exponent of photon's circular orbit.

\section{Summary}\label{sec6}

In this paper, we investigated the QNMs of the massive scalar field with a nonminimal coupling in the higher-dimensional near Nariai-type extremal rotating dS black hole. Particularly, we focused on the case of single rotation because there are various phase diagrams, as shown in Fig.\,\ref{fig:fig1}. Further, we expected to show how the quasinormal frequencies are related to the mass and nonminimal coupling of the scalar field. To obtain the quasinormal frequency, we started from the Lagrangian of the massive scalar field with a nonminimal coupling. Then, satisfying the field equation, the solutions at the boundaries provided the conditions for QNMs. According to the phase diagrams, the black hole with a single rotation can be near Nariai-type extremal in any number of dimensions. Therefore, under the conditions of QNMs, the radial equation of the scalar field was solved in the near Nariai-type extremal black hole. In the tortoise coordinate, the potential term for the Schr\"{o}dinger-like radiation equation is the P\"{o}shl-Teller potential. Under the boundary conditions of the QNMs and the P\"{o}shl-Teller potential, we could obtain the generalized form of the quasinormal frequency in $D$ dimensions. The real part of the frequency included contributions of mass and nonminimal coupling of the scalar field. This can be only discussed in this analysis according to the QNMs. By the Lyapunov exponent, this cannot be seen because there is no correspondence of the nonminimal coupling to a particle. Further, the imaginary part of the frequency implies the decay rate of the QNMs given in simple form. Particularly, the dominant QNM is found to be in the half of the surface gravity for a given dimensionality. The decay rate of the dominant QNM is now known as a close relationship to the SCC conjecture. Here, in the four dimensions, $\beta<1/2$ for the Kerr-dS black hole. Then, the scalar field is inextendible across the Cauchy horizon, and its energy is divergent. Thus, the SCC conjecture is still valid for a near Nariai-type extremal dS black hole with a single rotation in four and higher dimensions. This validity is consistent with previous studies.

\vspace{10pt} 

\noindent{\bf Acknowledgments}

\noindent This work was supported by the National Research Foundation of Korea (NRF) grant funded by the Korea government (MSIT) (NRF-2018R1C1B6004349) and the Dongguk University Research Fund of 2019.

\end{document}